\documentclass[twocolumn,showpacs,preprintnumbers,amsmath,amssymb]{revtex4}

\usepackage{graphicx}
\usepackage{dcolumn}
\usepackage{epsfig}

\begin{document}

\title{Dynamic phase separation of fluid membranes with rigid
  inclusions}

\author{Thomas R.\ Weikl} \affiliation{Max-Planck-Institut f\"ur
  Kolloid- und Grenzfl\"achenforschung, 14424 Potsdam, Germany}
\email{Thomas.Weikl@mpikg-golm.mpg.de}

\date{\today}

\begin{abstract}
  Membrane shape fluctuations induce attractive interactions between
  rigid inclusions. Previous analytical studies showed that the
  fluctua\-tion--\-induced pair interactions are rather small compared
  to thermal energies, but also that multi-body interactions cannot be
  neglected. In this article, it is shown numerically that shape
  fluctuations indeed lead to the dynamic separation of the membrane
  into phases with different inclusion concentrations. The tendency of
  lateral phase separation strongly increases with the inclusion size.
  Large inclusions aggregate at very small inclusion concentrations
  and for relatively small values of the inclusions' elastic modulus.
\end{abstract}


\maketitle

\section{Introduction}

Biological and biomimetic membranes consist of a lipid bilayer with
various types of macromolecules such as proteins
\cite{alberts94,lipowsky95}. Many of these macromolecules are
incorporated in the bilayer, others are covalently bound or adsorbed
to the membrane. The membranes are fluid and often tend to
phase-separate and to form domains or `rafts' with different molecular
composition. In biological membranes, the presence of domains has been
linked to specific functions in signaling \cite{simons97}, budding
\cite{schekman96}, or cell adhesion \cite{monks98,grakoui99}. In some
cases, the domain formation is caused by a separation of the lipid
bilayer into phases with different lipid composition
\cite{keller98,dietrich01}.  In other cases, the phase separation of
the membrane appears to be driven by attractive interactions between
membrane inclusions \cite{sackmannGroup}.

Besides direct interactions such as van der Waals or electrostatic
forces, membrane inclusions are subject to indirect interactions which
are mediated by the membrane.  Some of these membrane-mediated
interactions are {\em static}, i.e.~they arise from local
perturbations of the bilayer structure or shape around the inclusions.
Trans--membrane proteins which exhibit a hydrophobic mismatch with
respect to the lipid bilayer cause a perturbation of the bilayer
thickness. This thickness perturbation has been found to induce
attractive interactions between the proteins
\cite{dan94,fournier98,may99,harroun99}. Similar interactions due to
membrane thickness perturbations have also been proposed for adsorbed
particles \cite{schiller00}.  Membrane inclusions with conical shape
\cite{goulian93,tw98,kim98_99,sintes98,dommersnes99a,dommersnes99b,biscari02} 
or membrane--anchored polymers \cite{breidenich00,bickel00_01} cause
local perturbations of the membrane curvature which induces attractive
or repulsive interactions.

\begin{figure}[b]
  \epsfig{figure=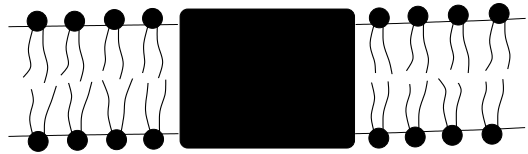,
    width=0.8\linewidth}
\caption{Rigid inclusion in the lipid bilayer \label{inclusions}}
\end{figure}

Other indirect interactions are {\em dynamic}, i.e.~they are mediated
by shape fluctuations of the membrane. In this article, rigid membrane
inclusions are considered which interact due to the suppression of
membrane shape fluctuations
\cite{goulian93,netz95,park96,golestanian96,netz97,holzloehner99,tw01a,helfrich01},
see Fig.~\ref{inclusions}.  Fluctuation--induced interactions have
also been found for specific receptors or stickers which locally bind
opposing membranes during adhesion \cite{bruinsma94,tw00,tw01b}.

The fluctuation--induced pair interactions of rigid membrane
inclusions have been studied intensively
\cite{goulian93,dommersnes99a,dommersnes99b,netz95,park96,golestanian96,%
  netz97,helfrich01}. The pair interaction of rigid disks with radius
$r$ and distance $L$ was found to be
\begin{equation}
G(L) = - 6 k_B T  (r/L)^4 + O[(r/L)^6] \label{pi}
\end{equation}
in the absence of an external membrane potential and lateral tension
\cite{goulian93,golestanian96}. Here, $O[(r/L)^6]$ stands for terms of
sixth or higher order in $r/L$.  The prefactor of the leading term in
Eq.~(\ref{pi}) was first 12, and later corrected to 6
\cite{golestanian96}, see also \cite{dommersnes99a,dommersnes99b,park96,helfrich01}.
Since the distance $L$ has to be larger than the inclusion diameter $2
r$, this term is only a fraction of the thermal energy $k_B T$.
However, higher--order terms in $(r/L)$ might be relevant at small
inclusion separations and thus contribute to the phase behavior of a
membrane with many inclusions. Deducing the phase behavior of such a
membrane is also complicated by non-trivial multi-body interactions
\cite{netz97}.

In this article, it is shown numerically that membrane--shape
fluctuations indeed lead to the aggregation of rigid inclusions.  The
phase behavior of a discretized membrane with rigid inclusions is
considered in Monte Carlo simulations. The membrane consists of
quadratic patches with linear extension $a$ which corresponds to the
smallest possible wavelength for bending deformations. Computer
simulations of molecular membrane models indicate that this length
scale is about 6 nm for a lipid bilayer with a thickness of about 4 nm
\cite{goetz99}.  Above a critical value $K^*$ for the stiffness
modulus of the inclusions, the membrane is found to separate into an
inclusion--rich and an inclusion--poor phase.  The aggregation
tendency of the inclusions strongly increases with the size $Q$, which
is reflected in a decrease of the critical stiffness $K^*$ with the
inclusion size. Large inclusions also aggregate already at relatively
small inclusion concentrations. Here, quadratic inclusions with a size
$Q$ of $2\times 2$, $3\times 3$, or $4\times 4$ membrane patches are
considered, which extends previous results on smaller inclusions with
the size of a single patch \cite{tw01a}.  For these inclusions, the
critical stiffness is found to decrease according to the power law
$K^*\sim Q^c$ with exponent $c\simeq -0.70$.

\section{General model}

\begin{figure}
\begin{center}
  \epsfig{figure=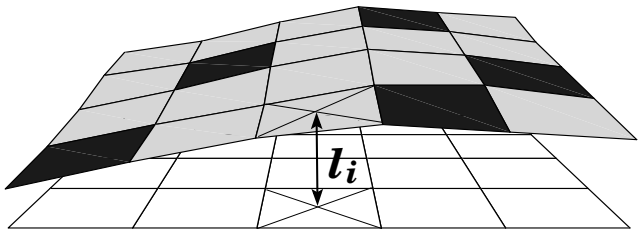, width=\linewidth}
\end{center}
\caption{A membrane segment containing inclusions with the size of one
  lattice site. The segment consists of $5\times 4$ membrane patches
  which are labeled by the lattice sites $i$.  Membrane patches with
  black inclusions correspond to occupation numbers $n_i=1$, while
  grey patches without inclusion have occupation numbers $n_i=0$.  The
  local deviation of the membrane from the white reference plane is
  denoted by $l_i$.
\label{model}}
\end{figure}

The configurations of a fluctuating membrane with inclusions can be
described by a field $l$ for the membrane shape and a concentration
field $n$ for the inclusions \cite{lipowsky96}. For a membrane which
is on average planar, the membrane shape is usually given by the
deviation $l(x,y)$ from a reference plane with coordinates $x$ and
$y$.  Here, we discretize the reference plane into a square lattice
with lattice constant $a$ which corresponds to the smallest possible
wavelength for bending deformations.  The inclusion positions are then
given by occupation numbers $n_i=0$ or $1$ where $n_i=1$ denotes the
presence of an inclusion at the lattice site $i$ of the reference
plane, see Fig.~\ref{model}.

In the absence of inclusions, the discretized bending energy per
lattice site can be written as
\begin{equation}
{\cal E}_i^{M}(l) = {\textstyle\frac{1}{2}} a^2 \kappa_o(c_{i,x}+c_{i,y})^2
\label{Emem}
\end{equation}
where $\kappa_o$ is the bending rigidity of the lipid bilayer, and
$\frac{1}{2}(c_{i,x}+c_{i,y})$ is the local mean curvature of the
membrane \cite{helfrich73}. Here,
\begin{eqnarray}
c_{i,x}=(l_{x+a,y}+l_{x-a,y}-2 l_{x,y})/a^2\\
c_{i,y}=(l_{x,y+a}+l_{x,y-a}-2 l_{x,y})/a^2  
\end{eqnarray}
are the discretized curvatures in $x$- and $y$-direction at the
lattice site $i$ with coordinates $(x,y)$.
The rigid inclusions here are characterized by the elastic 
energy per site
\begin{equation}
{\cal E}_i^{I}(l) =  {\textstyle\frac{1}{2}} a^2 
K (c_{i,x}^2 + c_{i,y}^2) \label{inc}
\end{equation}
with the stiffness modulus $K$ \cite{footnote}. For $K\to\infty$, such
inclusions are completely rigid and suppress any local curvature at
the inclusion position similar to the rigid disks or rods studied in
\cite{goulian93,park96,golestanian96,helfrich01}. In contrast,
inclusions with increased bending rigidity as considered in
\cite{netz95,netz97,tw01a} only suppress fluctuations of the total
curvature $c_{i,x} + c_{i,y}$, but not saddle-type fluctuations with
$c_{i,x}=- c_{i,y}$.

The grand-canonical Hamiltonian of a membrane containing inclusions
with the size of one lattice site can be written as
\begin{eqnarray}
{\cal H}_{Q=1}\{l,n\} &=& \sum_i \Big\{
(1-n_i) {\cal E}_i^{M}(l) 
+ n_i\left[{\cal E}_i^{I}(l)-\mu\right]\nonumber\\
&& \hspace*{1cm}+\, V(l_i)\Big\}
\end{eqnarray}
were $\mu$ is the relative chemical potential of the inclusions, and
$V(l_i)$ is the external membrane potential.  On lattice sites with
occupation numbers $n_i=1$ indicating the presence of inclusions, the
elastic energy is given by ${\cal E}_i^{I}(l)$. On lattice sites with
$n_i=0$, the elastic energy is the energy ${\cal E}_i^{M}(l)$ of the
lipid bilayer.

\begin{figure}
\begin{center}
  \epsfig{figure=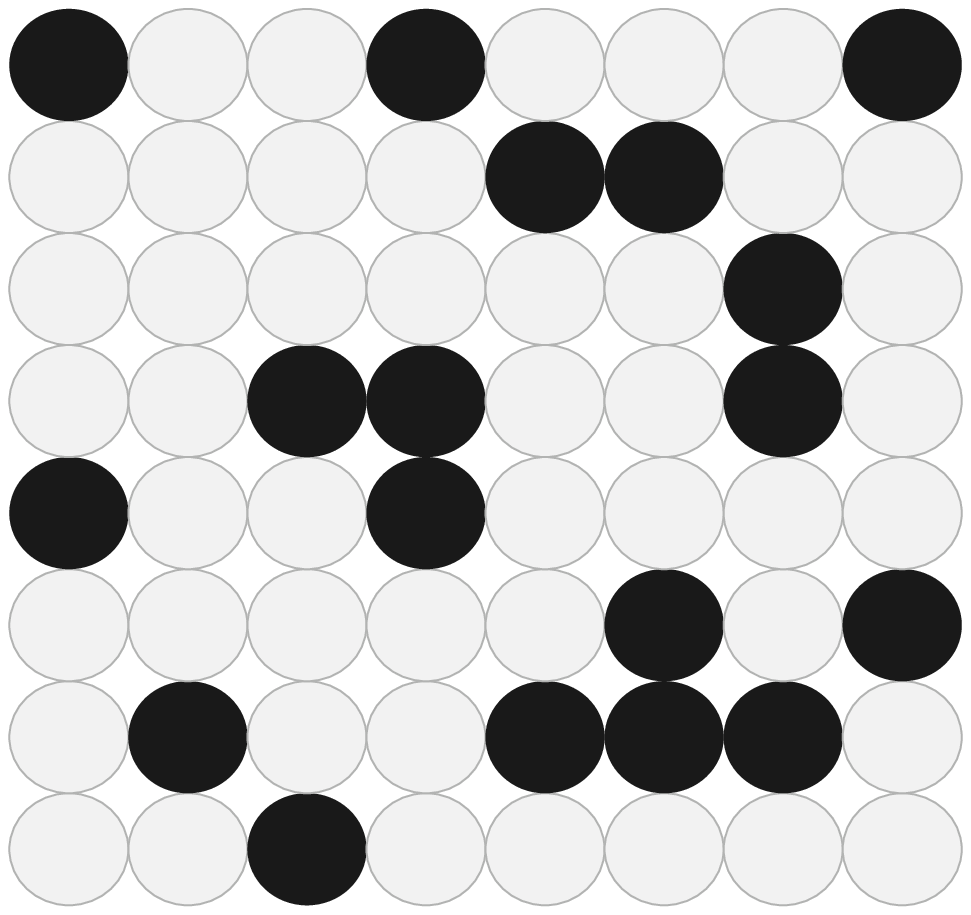, width = 0.4\linewidth}
  \hspace{0.7cm} 
  \epsfig{figure=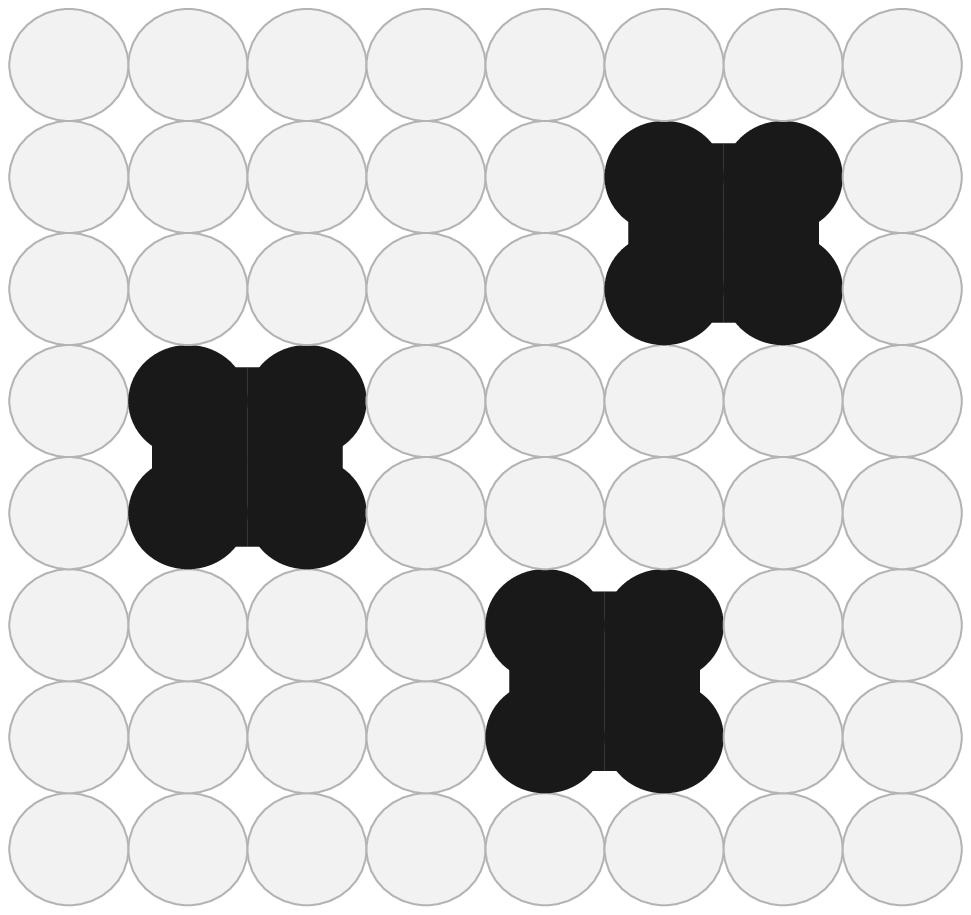, width = 0.4\linewidth}
  
  \vspace{0.5cm}
  
  \epsfig{figure=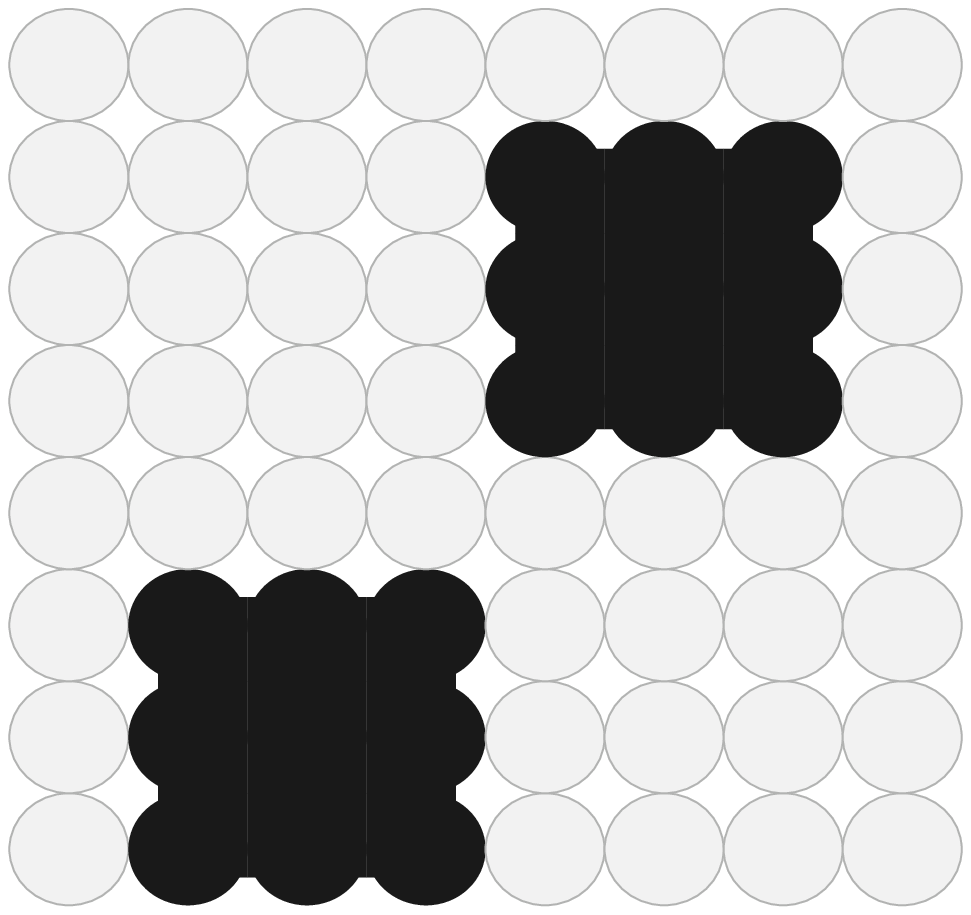, width = 0.4\linewidth}
  \hspace{0.7cm} 
  \epsfig{figure=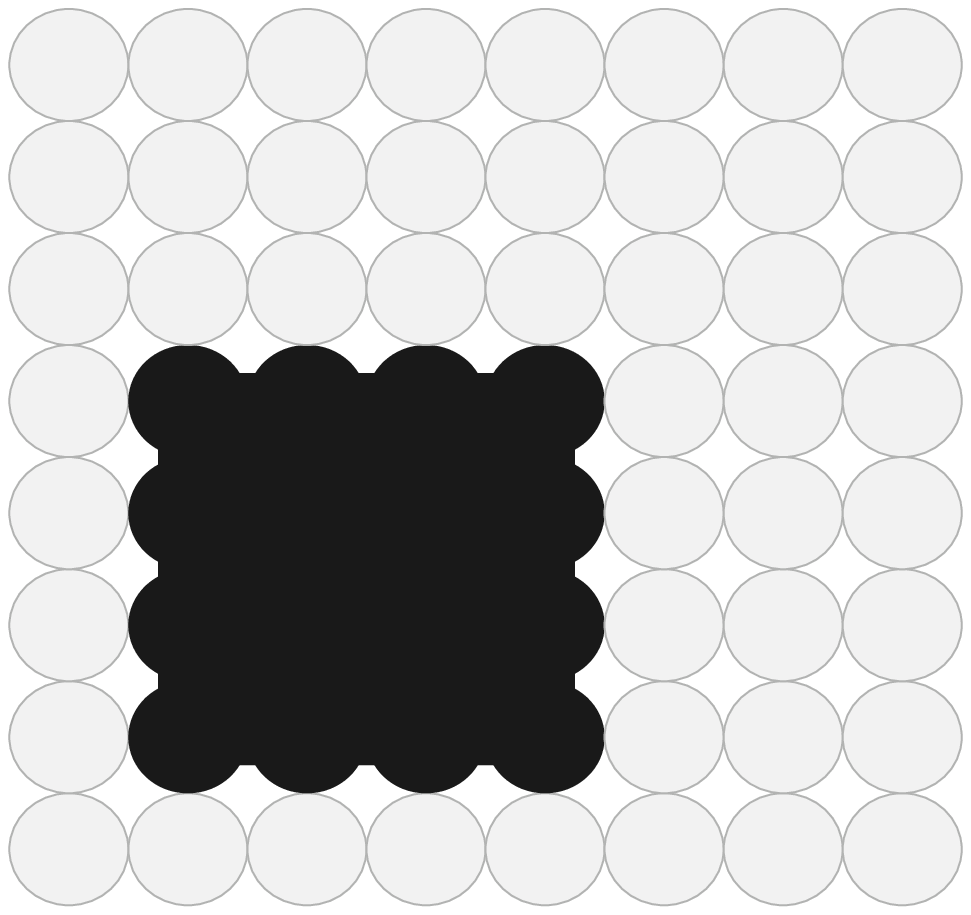, width = 0.4\linewidth}
\end{center}
\caption{Top view of membrane segments with inclusions of
  size $Q=1$, $2\times 2$, $3\times 3$, and $4\times 4$, respectively.
  The inclusions are shown in black.
\label{inclusionsII}}
\end{figure}

The larger inclusions considered here have an area of $Q=2\times
2$, $3\times 3$, or $4 \times 4$ membrane patches or lattice sites,
see Fig.~\ref{inclusionsII}. The elastic energy at every lattice site
of an inclusion is given by Eq.~(\ref{inc}). The larger inclusions
thus can be seen as quadratic arrays of small inclusions with the size
of one lattice site. The grand--canonical Hamiltonian for a membrane
with larger inclusions can be formally written as \cite{tw00}
\begin{eqnarray}
{\cal H}_{Q}\{l,n\}&\!=\!&\sum_i\bigg\{ {\cal E}_i^{M}(l) +
    n_i\bigg[\sum_{q=1}^Q \Big({\cal E}_{iq}^I(l) 
    - {\cal E}_{iq}^M(l)\Big)-\mu\bigg] \nonumber\\
&& \hspace{1cm} +\, V(l_i) \bigg\} + \sum_{\langle ij\rangle}
W_{ij} n_i n_j \label{HQ}
\end{eqnarray}
where $\{i1,\ldots,iQ\}$ denotes quadratic arrays of $Q=2\times 2$,
$3\times 3$, or $4\times 4$ lattice sites.  The position of an
inclusion given by $n_i=1$ corresponds to one of the lattice sites
occupied by the inclusion, e.g.~the center of an inclusion with size
$Q=3\times 3$.  The hard-square interaction
\begin{eqnarray}
W_{ij} &=& \infty \hspace{0.3cm} \mbox{for} \hspace{0.2cm} j
\hspace{0.2cm} \mbox{in} \hspace{0.2cm}  A_i^Q \nonumber\\ 
&=& 0 \hspace{0.5cm} \mbox{otherwise} \label{hardsquare}
\end{eqnarray}
prevents any overlap of inclusions. Here, $A_i^Q$ denotes the
exclusion area of an individual inclusion with size $Q$ at lattice
site $i$.

In the following, the external potential of the membrane is taken to
be the harmonic potential
\begin{equation}
V(l_i) = m l_i^2/2  \label{harmpot}
\end{equation}
with potential strength $m$.  The harmonic potential introduces an
additional length scale, the correlation length
$\xi=(4a^2\kappa_o/m)^{1/4}$ for the deviation field $l$, see, e.g.,
Ref.~\cite{netz95}.  Membrane fluctuations on length scales larger
than the correlation length $\xi$ are suppressed by the harmonic
potential, while fluctuations on smaller scales are governed
predominantly by the elastic energy of the membrane.

The membrane model defined by the Eqs.~(\ref{Emem}) to (\ref{HQ}) has
four characteristic dimensionless parameters, as can be shown be
introducing the rescaled deviation field
\begin{equation}
z\equiv (l/a)\sqrt{\kappa_o/(k_B T)}
\end{equation}
These parameters are the ratio $K/\kappa_o$ of the inclusion modulus
and the bare membrane rigidity, the dimensionless chemical potential
$\mu/(k_B T)$ for the inclusions, the rescaled potential strength
$\tilde{m}\equiv m a^2/\kappa_o$, and the inclusion size $Q$.

\section{Monte Carlo simulations}
%

\begin{figure}[!b]
\begin{center}
  \epsfig{figure=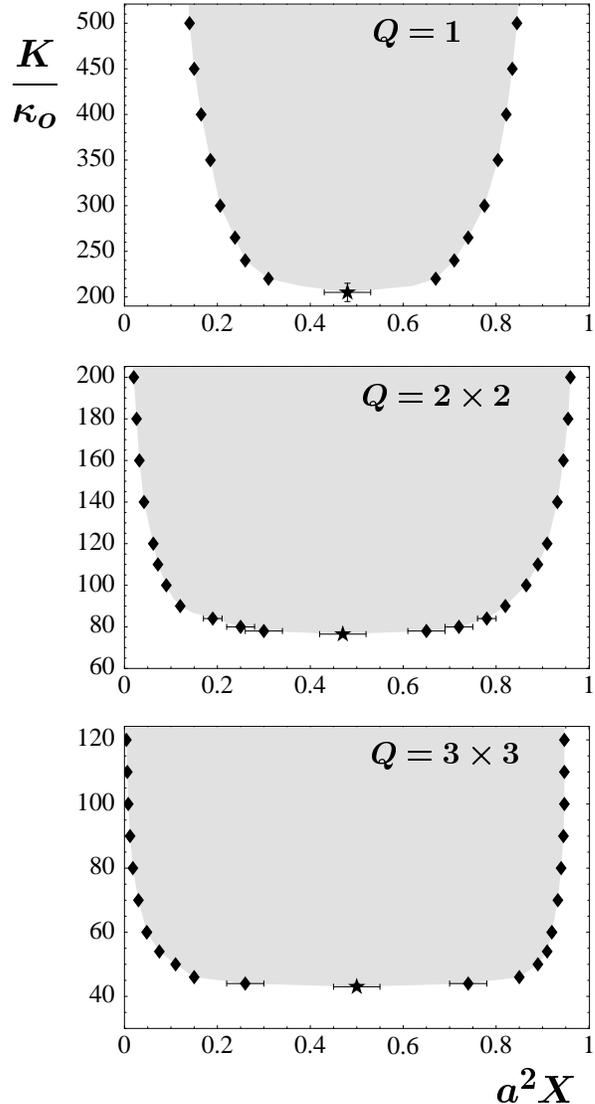,
    width=0.9\linewidth}
\end{center}
\caption{Phase diagrams for inclusions with size $Q=1$, $Q=2\times 2$,
  and $Q=3\times 3$ as function of the inclusion concentration $X$ and
  stiffness $K$ in units of the bare membrane rigidity $\kappa_o$.
  The rescaled potential strength is $\tilde{m}=0.01$. Inside the
  shaded coexistence regions, the membrane separates into an
  inclusion--rich and an inclusion--poor phase with concentrations
  given by the lines of Monte Carlo points.  The critical points are
  represented by stars.
\label{phaseDiagrams}}
\end{figure}

To deduce the phase behavior from Monte Carlo simulations, the
inclusion concentration $X\equiv Q\langle n_i\rangle/a^2$ is
determined as a function of the chemical potential $\mu$ for various
values of the inclusion stiffness $K$, size $Q$, and the rescaled
potential strength $\tilde{m}$.  A first-order phase transition is
reflected in a discontinuity of $X(\mu)$ at a certain chemical
potential $\mu_{tr}$.  The two limiting values of $X(\mu)$ at
$\mu_{tr}$ are the inclusion concentrations of the coexisting phases,
an inclusion-rich and an inclusion-poor phase.  To determine the
inclusion concentration $X$ at a given value of $\mu$, Monte Carlo
simulations are performed with up to $10^7$ Monte Carlo steps per
lattice site on a lattice with $120\times 120$ sites and periodic
boundary conditions. Each Monte Carlo step consists in attempted local
moves of the rescaled deviation field $z$ and of the concentration
field $n$ on all lattice sites.
For rescaled potential strengths $\tilde{m}\ge 0.01$ as considered
here, the correlation length of the membrane is much smaller than the
lateral extension of the lattice, and finite size effects are
negligible.

In Fig.~\ref{phaseDiagrams}, phase diagrams as a function of the
inclusion modulus $K$ are shown for inclusions with size $Q=1$,
$2\times 2$, and $3\times 3$. The rescaled potential strength is
$\tilde{m}=0.01$.  At points $(X,K)$ inside the shaded 2-phase
coexistence regions, the membrane separates into an inclusion-rich and
an inclusion-poor phase.  The inclusion concentrations in the
coexisting phases are given by the lines of Monte Carlo data points,
the critical points are represented by stars.  The extent of the
2-phase regions strongly increases with the inclusion size $Q$, wich
is reflected (i) in a strong decrease of the critical stiffness $K^*$ and
(ii) in an increase of the width of the coexistence region with the
inclusion size. For inclusions with size $Q=3\times 3$, the membrane
phase-separates already at very small inclusion concentrations.

\begin{figure}
\begin{center}
  \epsfig{figure=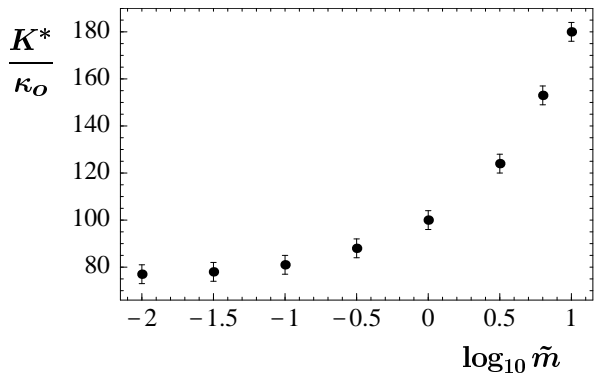, width = 0.9\linewidth}
\end{center}
\caption{Critical inclusion stiffness $K^*$ 
  as a function of the rescaled potential strength $\tilde{m}$ for
  inclusions with size $Q=2\times 2$.  The critical stiffness $K^*$ is
  given in units of the bare membrane rigidity $\kappa_o$ and
  increases with $\tilde{m}$ since the harmonic potential suppresses
  membrane fluctuations.
\label{Q2x2}}
\end{figure}

\begin{table}
\caption{Critical values $K^*/\kappa_o$ of the inclusions stiffness
$K$ in units of the bare membrane rigidity $\kappa_o$ for 
various inclusions sizes $Q$ and rescaled potential strengths
$\tilde{m}$. Values for $K^*/\kappa_o$ in the limit 
$\tilde{m}\to 0$ are obtained by extrapolation from the critical values 
at finite $\tilde{m}$ shown in the columns 2 to 4. 
\label{Kstar}}
\begin{ruledtabular}
\begin{tabular}{c|ccc|c}
& $\tilde{m}=1$ & $\tilde{m}=0.1$ & $\tilde{m}=0.01$ & $\tilde{m}\to 0$\\ 
\hline $Q=1$ & $300\pm 10\;$ & $225\pm 10$ &  
$205\pm 10$ & $198\pm 10$ 
\\ $Q=2\times 2$ & $100\pm 2$ & $81\pm 2$ &  $77\pm 2$ &
$76\pm 2 $ \\ $Q=3\times 3$ & $55\pm 2$ & $44\pm 2$ &  $43\pm 2$ &
$42.9\pm 2 $ \\ $Q=4\times 4$ & $37\pm 2$ & $29\pm 2$ &  $28\pm 2$ &
$27.9\pm 2 $
\end{tabular}
\end{ruledtabular}
\end{table}

The external harmonic potential (\ref{harmpot}) suppresses membrane
shape fluctuations on length scales larger than the correlation length
$\zeta=a(4/\tilde{m})^{1/4}$. Since the phase separation of the
membrane is driven by the fluctuations, the critical stiffness $K^*$
increases with the rescaled potential strength $\tilde{m}$, see
Fig.~\ref{Q2x2}.  For weak potentials with small values of
$\tilde{m}$, the critical stiffness is rather independent of
$\tilde{m}$ and tends towards a limiting value, since the correlation
length $\zeta$ then is much larger than the average distance between
neighboring inclusions.

\begin{figure}
\begin{center}
  \epsfig{figure=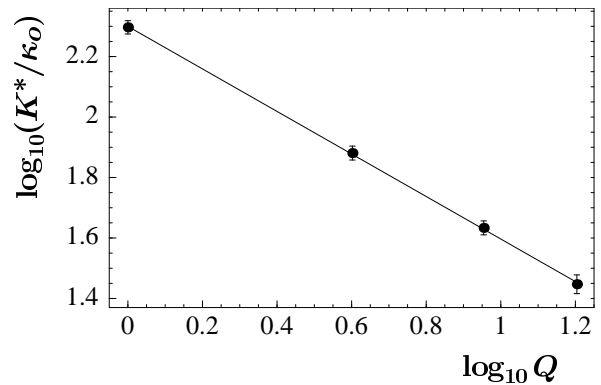, width = 0.9\linewidth}
\end{center}
\caption{Scaling plot for the critical inclusion stiffness $K^*$ as  a
  function of the size $Q$ in the limit of the rescaled potential
  strength $\tilde{m}\to 0$, see also Table 1.  The straight line has
  the slope $-0.70$ which corresponds to the exponent $c$ in
  Eq.~(\ref{powerlaw}).
\label{m0}}
\end{figure}

In the absence of an external potential, i.e.~for $\tilde{m}=0$, the
critical stiffness $K^*$ of the inclusions only depends on the
inclusion size $Q$.  Since the increasing correlation length $\zeta$
leads to finite size effects in Monte Carlo simulations with
$\tilde{m}=0$, the critical stiffness $K^*$ for $\tilde{m}\to 0$ is
determined here by extrapolation, see Table 1.  For the inclusion
sizes considered in this article, the functional dependence of the
critical values $K^*$ for $\tilde{m}\to 0$ on the size $Q$ can be
approximated by the power law
\begin{equation}
  K^* \sim Q^c \label{powerlaw}
\end{equation}
with exponent $c= -0.70\pm 0.01$, see Fig.~\ref{m0}.

\section{Scaling analysis}

Large inclusions with size $Q>1$ can be seen as quadratic arrays of
small inclusions with the size of one lattice site.  These inclusions
aggregate at much smaller values of the stiffness $K$ than inclusions
with size $Q=1$, see above. In order to understand this behavior, it
is instructive to consider the free energy difference $\Delta F$
between the two membrane states: The uniform state in which the
inclusions are more or less homogeneously distributed throughout the
membrane, and the phase-separated state in which the inclusions are
aggregated. If $\Delta F$ is negative, the membrane will be in its
homogenous state; if $\Delta F$ is positive, the membrane will
phase-separate. For a given area concentration $X$ of the inclusions, 
the free energy difference between the uniform and the aggregated state 
can be written as
\begin{equation}
\Delta F(K,Q) = \Delta F_{int}(K,Q) - T \Delta S_{mix}(Q) \label{fkq}
\end{equation}
in the absence of an external potential, i.e.~for $m=0$. Here, $\Delta
F_{int}(K,Q)$ is the difference in the dynamic interaction free energy
of the inclusions, which is induced by the shape fluctuations of the
membrane, and $\Delta S_{mix}(K,Q)$ is the difference in the entropy
of mixing.

The second term in Eq.~(\ref{fkq}) is dominated by the entropy of
mixing in the homogeneous state. For small area concentrations
$X=Q\langle n_i\rangle/a^2$ of the inclusions, this entropy
of mixing is proportional to the
number of inclusions, and 
$\Delta S_{mix}$
scales as
\begin{equation}
\Delta S_{mix}(Q) \sim 1/Q  \label{smix}
\end{equation}
Eq.~(\ref{smix}) simply results from the fact
that the number of inclusions is proportional to $1/Q$ for given area
concentration $X$ of the inclusions.

The term $\Delta F_{int}(K,Q)$ is dominated by the interaction free
energy of the inclusions in the aggregated state. For $Q>1$, the
inclusions are rather densely packed in this state with an area
fraction $a^2 X$ larger than 0.9 (see Fig.~\ref{phaseDiagrams}), and
have contact with neighboring inclusions almost along the whole
circumference of length $4 \sqrt{Q}$.
If one assumes that the interaction free energy in the aggregated
state is proportional to the ratio of the inclusion circumference $4
\sqrt{Q}$ and the area $Q$, the scaling form of the interaction free
energy difference can be estimated as
\begin{equation}
\Delta F_{int}(K,Q) \simeq f(K)/\sqrt{Q}  \sim K^m/\sqrt{Q} \label{dfint}
\end{equation}
presupposing powerlaw-form for $f(K)$. According to this estimate, the
interaction free energy $\Delta F_{int}$ decreases with the inclusion
size $Q$ proportional to $Q^{-1/2}$, following the decrease of the
`surface-to-area' ratio of the inclusions. The critical stiffness
$K^*$ obtained from $\Delta F=0$ then scales as
\begin{equation}
K^* \sim Q^{-1/(2m)}
\end{equation}
Comparing with the exponent $c\simeq -0.70$ from the Monte Carlo
simulations (see Fig.~\ref{m0}), one obtains
\begin{equation}
m = -1/(2c)\simeq 0.71 
\end{equation}
Thus, for the stiffness values $K$ and inclusion sizes $Q$ considered
here, the dynamic interaction free energy $\Delta F_{int}$ of the
inclusions appears to increase proportional to $K^{0.71}$.

\section{Discussion}

In summary, I have considered the dynamic phase behavior of a
discretized membrane with rigid inclusions using Monte Carlo
simulations. The phase behavior strongly depends on the inclusion
size. For inclusion sizes ranging from $Q=1$ to $Q=4\times 4$ lattice
sites or membrane patches, the critical stiffness $K^*$ decreases with
the size as $K^*\sim Q^{-0.70}$ in the absence of an external
potential, see Fig.~\ref{m0}.  The lateral extension of a membrane
patch, the lattice spacing $a$, corresponds to the cut-off length for
membrane shape fluctuations, which has been estimated as 6 nm for a
membrane with a thickness of about 4 nm \cite{goetz99}.  In biological
or biomimetic membranes, rigid objects with an extension larger than 6
nm may correspond to large trans-membrane proteins, aggregates of
proteins and other macromolecules, or, more general, membrane domains
with increased elastic moduli.  Colloidal particles adsorbed on
membranes suppress membrane fluctuations similar to rigid inclusions.
In general, membrane inclusions and membrane-adsorbed particles may 
have a variety of shapes and, therefore, orientational degrees of freedom
\cite{dommersnes99b,golestanian96,holzloehner99}.
Here, I have only considered quadratic objects on a square lattice.
At high area concentrations $a^2 X > 0.8$, the phase behavior of the
membrane with quadratic inclusions of size $Q>1$ is complicated by the packing
transitions of the hard-square lattice gas \cite{runnels72,kinzel81}
which are not considered here.  These transitions are induced by the
hard-square interactions (\ref{hardsquare}) of the inclusions, but do
not depend on the inclusion stiffness in contrast to the dynamic phase
separation.

The inclusions considered here suppress fluctuations of the local
curvatures $c_{i,x}$ and $c_{i,y}$ in $x$- and $y$-direction at the
inclusion sites, see Eqs.~(3) to (5). In contrast, inclusions with
increased bending rigidity studied in \cite{netz95,netz97,tw01a} only
suppress fluctuations of the total curvature $c_{i,x}+c_{i,y}$, but
not `saddle-type' fluctuations with $c_{i,x}=-c_{i,y}$, which seems
somewhat less realistic. The phase behavior of these inclusions is
remarkably different from that of the rigid inclusions studied here.
Inclusions with increased bending rigidity do not interact in the
absence of an external membrane potential, since local fluctuations of
the total curvature at different membrane sites are not correlated in
the free membrane \cite{netz95}. Such correlations are only induced by
the external potential of the membrane. The fluctuation--induced
interactions between inclusions with increased bending rigidity attain
a maximum at a certain nonzero potential strength, but are always
considerably weaker than those of rigid inclusions characterized by
Eq.~(5), see Ref.~\cite{tw01a}. Intermediate cases between these
two types of inclusions can be studied by using an elastic energy
with two moduli for the inclusions \cite{tw01a}.

Rigid inclusions may also be subject to other membrane-mediated
interactions if they perturb the bilayer thickness or have a conical
or wedge-like shape, see introduction. In general, dynamic
fluctuation-induced interactions can be assumed to be additive to
static interactions arising from perturbations of the equilibrium
membrane structure, as long as these perturbations do not affect the
elastic moduli of the membrane.  The variety of membrane--mediated
indirect interactions often complicates the interpretation of
experimental results. The dynamic phase separation of a fluctuating
multi-component membrane in contact with a substrate has been recently
reported in \cite{marx02}. The membrane contains anchored PEG-polymers
and appears to phase separate into domains with different separation
from the substrate, which might result from different effective
bending rigidities for the domains since the fluctuation-induced
Helfrich repulsion \cite{helfrich78} between membrane and substrate
depends on the rigidity.  Fluctuation--induced interactions may also
contribute to the aggregation of latex spheres adsorbed to vesicles
reported in Ref.~\cite{koltover99}.  Lateral phase separation has also
been observed during the adhesion of biomimetic membranes with
specific receptors or stickers which bind to ligands in a supported
membrane \cite{sackmannGroup}.  Phase separation during membrane
adhesion may be induced by membrane fluctuations \cite{tw00,tw01b}, or
by an effective barrier in the interaction energy between the
membranes \cite{tw01b,komura00,bruinsma00}.  In the first case, the
aggregation tendency of the stickers strongly increases with the
sticker size, similar to the rigid inclusion considered here
\cite{tw00}.  The fluctuation--induced interactions between bound
stickers are also enhanced if the stickers are more rigid than the
surrounding membrane \cite{tw01b}.

\end{document}